\documentclass[prl,aps,twocolumn,draft,showpacs]{revtex4}
\usepackage[dvips]{graphicx}
\usepackage{graphpap}
\begin{document}





\title{Electronic shell effects and the stability of alkali nanowires}

\author{D.~F.~Urban$^a$, J.~B\"urki$^{a,b}$,  A.~I.~Yanson$^c$, I.~K.~Yanson$^d$, C.~A.~Stafford$^b$, J.~M.~van~Ruitenbeek$^e$, Hermann~Grabert$^a$}

\affiliation{${}^a$Physikalisches Institut,
Albert-Ludwigs-Universit\"at, D-79104 Freiburg, Germany \\
${}^b$Department of Physics, University of Arizona, Tucson, AZ 85721, USA\\
${}^c$Department of Physics, 510 Clark Hall, Cornell University, Ithaca, NY 14853, USA\\
${}^d$B.~Verkin Institute for Low Temperature Physics and Enginneering,
National Academy of Sciences, 310164 Kharkiv, Ukraine\\
${}^e$Universiteit Leiden, PO Box 9504, NL-2300 RA Leiden, The Netherlands}

\date{\today}

\begin{abstract}
Experimental conductance histograms for Na nanowires are analyzed
in detail and compared to recent theoretical results on the
stability of cylindrical and elliptical nanowires, using the
free-electron model. We find a one-to-one correspondence between
the peaks in the histograms and the most stable nanowire
geometries, indicating that several of the commonly observed
nanowires have elliptical cross sections.
\end{abstract}





\maketitle

%
%

\section{Introduction}

In the past eight years, experimental research on metal nanowires
has burgeoned, together with theoretical work describing these
nanoscale systems in which quantum effects play an important role
(see  Ref.~\cite{Agrait03} for a recent review). Metal nanowires
exhibit a number of interesting properties: their electrical
conductance is carried by a well-defined number of quantum
channels ~\cite{Agrait93,Brandbyge95,Krans95,Scheer98},
their shot-noise is suppressed by the Pauli
principle~\cite{Brom99}, and they are remarkably strong and
stable~\cite{Kondo97}. The attention to this research field is
accentuated by the ongoing miniaturization in the microelectronic
industry and the need to understand what modified properties may
be met upon further size reduction.

Crucial for answering the question of future applications is
knowledge about the stability of nanowires.
An important contribution to the energetics of metal nanowires comes
from the subband structure of the conduction electrons~\cite{Stafford97a,Ruitenbeek97,Yannouleas98}.
Indeed, a linear stability
analysis~\cite{Kassubek01,Zhang03,Urban03,Urban03b}
shows that electronic shell effects can stabilize and therefore
favor certain geometries.
The effects of shell filling on the abundance spectra of metal
clusters have been known for years. They explain the existence of
clusters with ``magic numbers,'' corresponding to full electronic
shells, which are observed more frequently than
others~\cite{Heer93}. More recently, electronic shell and
supershell structures have been reported for alkali metal
nanowires~\cite{Yanson99,Yanson00,Yanson01}. Theoretically, these
effects can be understood within a stability analysis including
wires with cylindrical and elliptical cross
sections~\cite{Urban03b}.

This paper is organized as follows: An overview on the theoretical
framework is presented in the next section, including the
discussion of the nanoscale free-electron model. Then experiments
measuring conductance histograms of alkali metal nanowires are
presented, followed by a more detailed analysis of the
experimental data. It is shown that theory allows prediction of
the shapes of wires produced in experiments. A short summary is
given in the last section.

%
%

\section{Theory}

One possibility to describe nanoscale contacts is via molecular
dynamics simulations~\cite{Landman90,Todorov96,Sorensen98} which
use short-ranged interatomic potentials suitable to describe the
bulk properties of metals. However, such an approach appears
problematic when applied to metal nanowires, in which
electron-shell effects~\cite{Yanson99,Yanson00} due to transverse
confinement are likely to be important. An alternative approach,
motivated by the success of the jellium approximation in
explaining the energetics of ultrasmall metal
clusters~\cite{Heer93,Brack93}, is a nanoscale generalization of
the free-electron model, developed by Stafford \emph{et~al.} \cite{Stafford97a}. This
approach allows for analytical calculations and was successfully
used in recent years to understand many properties of metal
nanowires.

\subsection{Nanoscale Free-Electron Model}
The free-electron model of a nanowire~\cite{Stafford97a}
treats the electrons as a non-interacting Fermi gas kept within
the wire by a confining potential, which is approximated in terms
of hard-wall boundary conditions. The discrete ions are replaced
by a constant positive jellium background charge. This model
requires good charge screening in the metal and an almost
spherical Fermi-surface. It is especially suitable for alkali
metals, but can also be applied to other monovalent metals such as gold.

A nanowire connecting two macroscopic electrodes is an open
quantum system, for which the Schr\"odinger equation is
most naturally formulated as a scattering problem. The fundamental
quantity describing the properties of the system is the
scattering matrix $S(E)$ connecting incoming and outgoing
asymptotic states of conduction electrons in the electrodes.
Transport properties can be expressed in terms of the submatrix $S_{12}$
describing transmission through the wire. For example,
the electrical conductance $G$ is given by the Landauer formula~\cite{Datta95}
\begin{eqnarray}
\label{gl::Conductance}
G&=&\frac{2e^2}{h}\int dE\,\frac{-\partial f(E)}{\partial
E}\;\mbox{Tr}\left\{S^{\dagger}_{12}(E)S_{12}(E)\right\},
\end{eqnarray}
where $f(E)$ is the Fermi-Dirac distribution function.
Similarly, thermodynamic properties can be expressed in terms of the
scattering matrix through the electronic density of states (DOS) $D(E)$
\begin{eqnarray}
\label{gl::DausS}
        D(E) &=&
    \frac{1}{2\pi i}\;\mbox{Tr}\left\{
    S^{\dagger}(E)\frac{\partial S}{\partial E} -
        \frac{\partial S^{\dagger}}{\partial E}S(E)\right\},
\end{eqnarray}
from which
the relevant thermodynamic potential for an open system, namely
the grand canonical potential $\Omega$, is obtained as
\begin{equation}
\label{gl::OmegaVonD}
    \Omega=-k_{B} T \int \!dE\; D(E) \;
    \ln\!\left[1+e^{-\frac{(E-\mu)}{k_{\mbox{\tiny B}}\,T}}
        \right],
\end{equation}
where $k_B$ is the Boltzmann constant, $T$ is the temperature and
$\mu$ is the chemical potential specified by the macroscopic
electrodes. Eqs.~(\ref{gl::Conductance}) through
(\ref{gl::OmegaVonD}) include a factor of 2 for spin degeneracy.

Many fundamental phenomena of metal nanowires have been
understood in terms of this nanoscale free-electron model. In
particular, theoretical results on conductance
quantization~\cite{Torres93,Buerki99a,Kassubek99},
cohesion~\cite{Stafford97a,Kassubek99} and current
noise~\cite{Buerki99b} are in good agreement with experiments. At
the mean-field level, electron-electron interactions can also be
included in the model in a straightforward
way~\cite{Kassubek99,Stafford99,Stafford00}.

In this approach, an atomic-scale contact between two macroscopic
electrodes can be considered as a waveguide for conduction
electrons (which are responsible for both electrical conduction
and cohesion in simple metals)~\cite{Stafford97a}. Each quantized
mode transmitted through the contact contributes one
quantum $G_0=2e^2/h$ to its conductance, and acts as a chemical bond, delocalized through the contact,
thus contributing to the cohesion of the wire.

\subsection{Stability and Shell Structure}
Metal nanowires are found to be remarkably stable, although one might expect
them to break up because of surface tension. Such an instability
arises from classical continuum mechanics and is known as the
Rayleigh instability~\cite{Cha81}. The key to the stability of
nanowires is quantum corrections to the classical stability analysis.

For wires aligned along the $z$-axis, the geometry is given by a
function $R(z,\varphi)$ when using cylindrical coordinates. The
linear stability of a nanowire with a given geometry can be analyzed by
calculating the energy change $\delta\Omega$ caused by a small
deformation $\delta R(z,\varphi)$ of the wire geometry. A stable
realization of a nanowire requires an increase of $\Omega$ for
all possible small deformations.
A constraint on the deformation arises from the fact that,
depending on material parameters, the deformed wire tries to find
a compromise between a volume conserving deformation and one
ensuring electroneutrality~\cite{Stafford99}, leading to the
constraint
\begin{equation}
\label{eq.constraint}
    {\cal N} \equiv k_F^3{\cal V} - \eta (3\pi k_F^2{\cal S}/8 )
= \rm{const.},
\end{equation}
where ${\cal V}$ is the volume of the wire, ${\cal S}$ its
surface area, and $k_F=2\pi/\lambda_F$ the Fermi wavevector of
the electrons. The parameter $\eta$ can be adjusted so as to fix
the value of the effective surface tension to the
material-specific value. In particular, $\eta=0$ corresponds to a
constant-volume constraint, and $\eta=1$ is the constraint of
constant Weyl charge. In this paper, the value $\eta=0.93$ is used,
corresponding to the surface tension of Na ($0.22\,\rm{N/m}$~\cite{Tyson77}.)

The stability was first analyzed for cylindrical geometries and
axisymmetric deformations, since these are the only destabilizing
deformations in the classical calculation. In a semiclassical
approach~\cite{Kassubek01,Zhang03}, the grand canonical potential
is divided into a smooth term depending on classical geometric
quantities (volume, surface, mean curvature), referred to as the Weyl
contribution~\cite{Brack97}, and a fluctuating quantum term. This
quantum potential is calculated using a Gutzwiller-type trace
formula~\cite{Gutzwiller90,Brack97}. One finds that for certain radii
electronic shell effects suppress the classical Rayleigh
instability at low temperatures. This leads to intervals of
stable (``magic'') radii, resulting in a series of stable
cylindrical wires with conductance
$G/G_0\in\{1, 3, 5, 6, 8, 10, 12, 14, 17, 23, \dots\}$.
Although all of those wires are predicted to be metastable, some of them
($G/G_0=5, 14, \dots$) are expected to be barely stable, and thus unlikely to be observable.
Since the semiclassical theory depends only trivially on the
length of the wire, a puzzling conclusion is that nanowires with
a magic radius remain stable whatever their length.

A full quantum calculation~\cite{Urban03} shows that
the suppression of the Rayleigh instability by electron shell
effects is supplemented by an interplay between the Rayleigh
and a novel Peierls-type instability missing in the semiclassical
approximation. In fact, this latter quantum mechanical
instability limits the maximal length of stable nanowires. For
lengths of order 10-1000 nm, the quantum calculation mainly
confirms the set of magic conductance values mentioned above, but
their corresponding intervals of stable radius values are reduced.
Wires with a conductance of $5G_0$ and $14G_0$ are unstable and
therefore missing.

Axial symmetry implies characteristic gaps in the sequence of
stable nanowires, which is not fully consistent with the
experimentally observed nearly perfect periodicity of the peak
positions as a function of $\sqrt{G}$.
The deviations can be accounted for neither by the
inclusion of disorder~\cite{Buerki01}, nor by the use of more
elaborate self-consistent jellium models~\cite{Puska01,Ogando02}.
Since gaps in the sequence of cylindrical nanowires arise from a
degeneracy of conductance channels, it is natural to assume that a
Jahn-Teller deformation breaking the symmetry can lead to more
stable deformed configurations.

In fact, the stability of nanowires with elliptic
cross-sections was recently examined theoretically~\cite{Urban03b} and stable
\emph{elliptical nanowires} were found. This extended
analysis confirms the stability of the cylindrical wires found in
the quantum calculation, even with respect to Jahn-Teller
distortions. This includes the fact that cylindrical wires with a
conductance of $5G_0$ and $14G_0$ are energetically unstable.
Secondly, a number of stable elliptical nanowires with an aspect
ratio $\varepsilon$ (which is defined as the ratio of the two
major semi-axes of the ellipse) greater than one are found. The
most stable elliptic wires have conductance values of
$G/G_0\in\{2,5,9,...\}$.

\begin{figure}
    \begin{center}
            \includegraphics[width=0.99\columnwidth,draft=false]{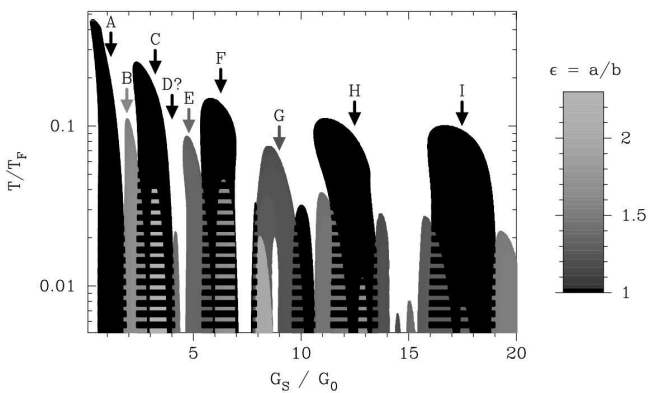}
    \end{center}
    \caption[]{\label{fig.stabdia} Calculated energetically stable cylindrical and elliptical nanowires
    as a function of temperature (in units of the Fermi temperature $T_F$).
    The surface tension is adjusted to a value of $0.22\,\rm{N/m}$ ($\eta=0.93$),
    corresponding to Na~\cite{Tyson77}.
    The aspect ratio $\varepsilon$ is coded via the gray scale, and
    hatched areas indicate multistability}
\end{figure}

Fig.~\ref{fig.stabdia} shows the corresponding stability diagram
in the low conductance range as a function of temperature. Shaded
areas indicate stability and the gray scale represents the value
of the aspect ratio $\varepsilon$. The $x$-axis is given by the
corrected Sharvin conductance $G_S$ of the wire~\cite{Agrait03}.
The elliptical wire with conductance
$9\,G_0$ is expected to be much more stable at low temperature than the neighboring
cylindrical wires at $8\,G_0$ and $10\,G_0$,
because it is stable up to a much higher temperature.

Extending this analysis to higher conductance values and
extracting the most stable configurations~\cite{Urban03b} leads to
a sequence of stable cylindrical and elliptical nanowires that
allows for a satisfying interpretation of the experiments on
shell and supershell structure~\cite{Yanson99,Yanson00,Yanson01}.

%
%

\section{Experiments}

Using remarkably simple experimental techniques, it is possible to
gently break a metallic contact and thus form conducting
nanowires. Besides the use of scanning tunneling microscopes
(STM), the mechanically controllable break junction (MCBJ) is the
most commonly used tool for the study of mechanical and transport
properties of metal nanowires and point contacts. By indenting one
electrode into another and then separating them, a stepwise
decrease in electrical conductance is observed, down to the
breakpoint when reaching a size of a single atom in cross
section. In addition to the conductance, the STM and MCBJ
techniques were used to study a variety of properties of metal
nanowires, i.e. cohesive force~\cite{Rubio96}, conductance
fluctuations~\cite{Ludoph00}, thermopower~\cite{Ludoph99}, and
shot noise~\cite{Brom99}. The highly reactive alkali metals have
been studied using a slightly modified MCBJ
technique~\cite{Krans95}.

\subsection{Conductance histograms}

\begin{figure}[b]
    \begin{center}
            \includegraphics[width=0.95\columnwidth,draft=false]{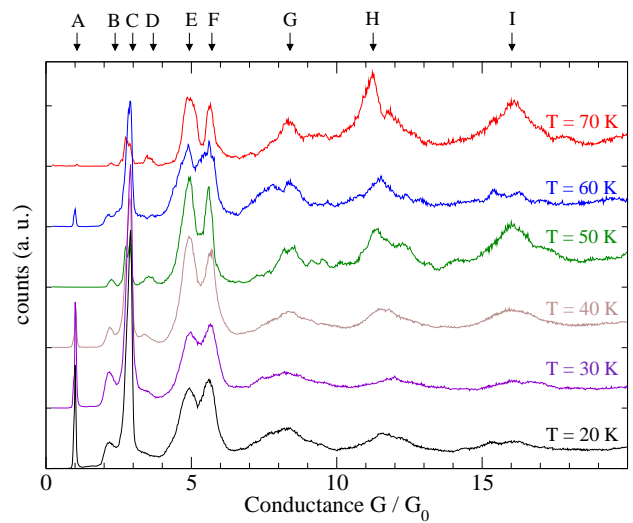}
    \end{center}
    \caption[]{\label{fig.histograms} Experimental conductance histograms for Na at different temperatures.
    The curves are vertically offset for clarity. The black arrows mark the approximate positions of
    the peaks that we consider in our analysis. (Data taken from Ref.~\cite{Yanson99}.)}
\end{figure}

Nanowires fabricated by the methods discussed above show a wide
variety of behaviours, with the atomic structure of the contacts
playing an important role. For example, each scan of the
dependence of conductance $G$ on the elongation $d$ is individual
in detail as the atomic configuration of each contact may be
widely different. Generally, the experiment doesn't allow for a
detailed knowledge of the atomic-scale structure of the contacts.
Nevertheless it is possible to extract information about possible
quantization effects by performing a statistical analysis. Many
individual conductance versus displacement curves together produce
a histogram of the probability for observing a given conductance
value, which is quite reproducible for a given metal and for
fixed experimental parameters.

An important subject for study is the alkali metals, since they are
nearly-free electron metals and therefore most closely approach
the predictions of the nanoscale free-electron model, described
in the previous section.
Fig.~\ref{fig.histograms} shows conductance histograms for sodium
recorded at different temperatures (taken from
Ref.~\cite{Yanson99}). The curves are vertically offset for
clarity and the amplitude has been normalized by the area under
each graph. Each histogram is constructed from 1000-2000
individual scans. A number of peaks with very different intensity
can be identified: Sharp peaks can be recognized at conductance
values near 1, 3, 5 and 6 $G_0$, while rather wide maxima are
found near 9, 12 and 17 $G_0$. In addition, there is a small peak
near $2\,G_0$ and a very weak structure near $4\,G_0$.
Results similar to those for Na have been obtained for Li and K.

%
%

\section{Further Analysis of Na Histograms}
\begin{figure}
    \begin{center}
            \includegraphics[width=0.95\columnwidth,draft=false]{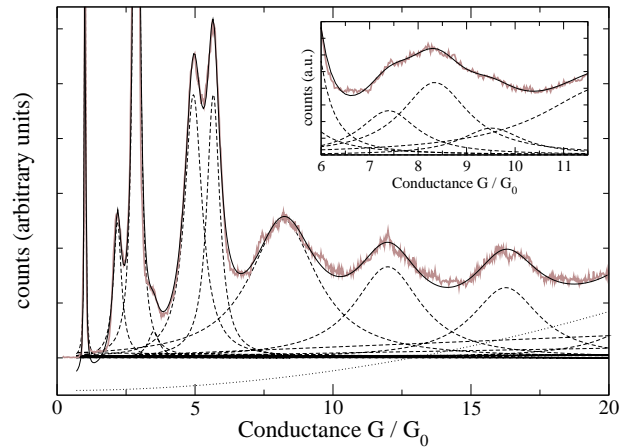}
    \end{center}
    \caption[]{Experimental conductance histogram data at $T=30K$ fitted with lorentzian peaks.
    The dashed lines show the individual peaks. Inset: A blow-up of the peak near $G=9\,G_0$
    fitted with two additional peaks of lower intensity is shown for comparison.
    \label{fig.lorentzfit}}
\end{figure}

In order to obtain more information about the stable geometries
that are realized in experiments, we have analyzed in detail the
conductance histograms for sodium recorded in the temperature
range of 20--70K (Fig.~\ref{fig.histograms}). We fitted each data
set with a function consisting of a sum of peaks plus a
background. The conductance peaks were taken to be of lorentzian
shape with height, width and peak position as fitting parameters.
We have chosen to take nine peaks, corresponding to those that are
clearly distinguishable in the data set. Their approximate
positions are marked with black arrows in
Fig.~\ref{fig.histograms} and they are labeled with capital
letters. This choice is not unique, as the fit can always be
improved by including additional broad peaks of lower intensities.
However, the parameters of additional peaks can not be determined
with high confidence, especially for noisy curves at higher
temperatures. Fig.~\ref{fig.lorentzfit} shows the fit for the data
at $T=30K$ and the inset shows a fit of the region near $9G_0$,
including two additional peaks of lower intensity, for comparison.

%
%

\begin{table}[b]
\begin{tabular}{|c|c|c|c|l|c|}
    \colrule
       peak &
       $\begin{array}{cc} \bar{G}_{\rm{exp.}} \\ \left[G_0\right]\end{array}$ &
       $\begin{array}{cc} \overline{\Delta G} \\ \left[G_0\right]\end{array}$ &
       $\begin{array}{cc} G_{\rm{th.}}  \\ \left[G_0\right]\end{array}$ &
       $\;\;\;\;$shape &
       $\begin{array}{cc}k_BT_{\rm{max}}^{\rm{(Na)}} \\ \left[eV\right]\end{array}$ \\
    \colrule
A &  1.0  & 0.1 & 1  &   cylindrical               & 1.50\\
B &  2.2  & 0.4 & 2  &   ellip. $\varepsilon=1.6$  & 0.36 \\
C &  2.8  & 0.2 & 3  &   cylindrical               & 0.81\\
D &  3.5  & 0.4 & 4  &   4-fold ?                  & ?\\
E &  4.9  & 0.6 & 5  &   ellip. $\varepsilon=1.3$  & 0.28\\
F &  5.6  & 0.5 & 6  &   cylindrical               & 0.48\\
G &  8.3  & 1.9 &
$\!\left\{\begin{array}{ccc}\!\!8\\\!\!9\\\!\!10\end{array}\right.$  &
$\!\!\!\begin{array}{lll}\,\mbox{cylindrical}\\\,\mbox{ellip.}\;\varepsilon=1.2\\\,\mbox{cylindrical}\end{array}$
&
$\begin{array}{ccc} 0.11 \\ 0.24 \\ 0.10 \end{array}$\\
H & 11.6   & 2.0 & 12  &   cylindrical  & 0.36\\
I & 15.9   & 2.4 & 17  &   cylindrical  & 0.32\\
    \colrule
\end{tabular}
\caption{ Experimental conductance, peak widths, theoretical conductance, and
          theoretical predicted shapes of the wires for
          the most important low
          conductance peaks. The right column lists the maximum temperature $T_{\rm{max}}$
          up to which the wires are expected to
          remain stable, which is related to the depth of the
          corresponding energetic minimum. Peak G is decomposed into three contributions as in the inset of Fig.\ \ref{fig.lorentzfit}.} \label{tab.wires}
\end{table}

The values of the fitting parameters now allow to analyze the
development and the contributions of the different conductance
peaks as a function of temperature. The peak positions stay almost
constant with maximum fluctuations of three percent within the
temperature range from 20--70K. All the peaks have a finite
(temperature dependent) width, and their positions are shifted
downward compared to the theoretical quantized conductance values
$G_{th}$. As an exception, we find peak B reproducibly at a
higher conductance value than expected. This is possibly due to
size constraints in building a nanowire of two channels by two
atoms, which for a short wire will have a rather large tunnelling
contribution of the third channel.

Table~\ref{tab.wires} lists average peak positions $\bar{G}_{exp}$,
average peak widths $\overline{\Delta G}$, and theoretical conductance
values $G_{th}$ in columns 2 to 4.
The last three columns sum up the results
of the theoretical stability calculations, taking into account the
most stable configurations only. The comparison with the
experimental data shows good agreement in that one can ascribe a quantized
conductance value to each experimental peak without ambiguity. In particular, it is
possible to predict the shape of the nanowires seen in
conductance histograms.

Peak D, with a conductance of $4 G_0$, is predicted to be barely stable in the present analysis.
However, one might expect to find more stable nanowires with cross sections
having a higher-order angular symmetry~\cite{Buerki04b}.
Such deformations cost significantly more surface energy, and can thus only be
energetically favorable at low conductance. Therefore they are not likely to modify
the stability diagram of Ref.~\cite{Urban03b}  significantly, but one might in particular
expect a wire with a four-fold angular symmetry to be stable at a conductance of $4 G_0$,
which could explain the observation of peak D.

\begin{figure}
    \begin{center}
            \includegraphics[width=0.95\columnwidth,draft=false]{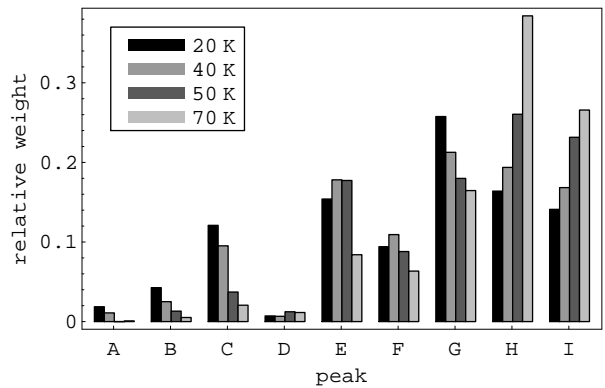}
    \end{center}
    \caption[]{Relative weights of the nine individual experimental conductance peaks (A to I) at four
    different temperatures, coded via the gray scale.\label{fig.weight}}
\end{figure}

The relative weight (intensity) of each peak in the experimental
histograms can also be extracted from the fit, and is shown in
Fig.~\ref{fig.weight} at four different temperatures. The
contribution of a stable wire to a histogram depends mostly on two
conditions: ({\sl i}) The wire has to be formed often enough to be
statistically relevant; ({\sl ii}) It needs to have a long enough
lifetime so as to be recorded. The low-temperature lifetime of a
nanowire is expected to increase with the maximum temperature up
to which it is predicted to remain linearly stable. This is
confirmed by recent calculations restricted to cylindrical
nanowires~\cite{Buerki04}.
Condition ({\sl i}) depends on the ability of the system to probe
new configurations, which is determined by the mobility of the
atoms and the density of stable geometries in configuration space.
Most of the very stable wires being cylindrical, one might expect
that highly deformed wires have a low probability to form. This
affects mainly the elliptic wire (B) with an aspect ratio of
$\varepsilon=1.6$. The elliptical wires (E) and (G) with a lower
aspect ratio have a remarkably higher weight. Interestingly, peak
(E) has a higher intensity than the neighbouring cylindrical peak
(F). The big relative weight of peak (G) is due to an overlay of
contributions of an elliptical wire with $G=9\,G_0$ and two
cylindrical wires with $G=8\,G_0$ and $G=10\,G_0$ as suggested
from the inset of Fig.~\ref{fig.lorentzfit}.
The decrease in the intensity of the low conductance peaks with
growing temperature might be explained by thermally induced
breaking of the contact.

%
%

\section{Conclusions}
We have presented a detailed analysis of experimental conductance
histograms for Na nanowires. Comparison with a theoretical
stability analysis of cylindrical and elliptical nanowires shows a
one-to-one relation between observed peaks and theoretically
stable geometries, and in particular allows a prediction of the
cross-sectional shape of the wires.

This research has been supported by the DFG through SFB 276 and
the EU Network DIENOW. JB and CAS acknowledge support from NSF
grant DMR0312028.

\bibliography{ssc_paper}

\begin{thebibliography}{41}
\expandafter\ifx\csname natexlab\endcsname\relax\def\natexlab#1{#1}\fi
\expandafter\ifx\csname bibnamefont\endcsname\relax
  \def\bibnamefont#1{#1}\fi
\expandafter\ifx\csname bibfnamefont\endcsname\relax
  \def\bibfnamefont#1{#1}\fi
\expandafter\ifx\csname citenamefont\endcsname\relax
  \def\citenamefont#1{#1}\fi
\expandafter\ifx\csname url\endcsname\relax
  \def\url#1{\texttt{#1}}\fi
\expandafter\ifx\csname urlprefix\endcsname\relax\def\urlprefix{URL }\fi
\providecommand{\bibinfo}[2]{#2}
\providecommand{\eprint}[2][]{\url{#2}}

\bibitem[{\citenamefont{Agra{\"\i}t et~al.}(2003)\citenamefont{Agra{\"\i}t,
  Levy~Yeyati, and van Ruitenbeek}}]{Agrait03}
\bibinfo{author}{\bibfnamefont{N.}~\bibnamefont{Agra{\"\i}t}},
  \bibinfo{author}{\bibfnamefont{A.}~\bibnamefont{Levy~Yeyati}},
  \bibnamefont{and} \bibinfo{author}{\bibfnamefont{J.~M.} \bibnamefont{van
  Ruitenbeek}}, \bibinfo{journal}{Phys. Rep.} \textbf{\bibinfo{volume}{377}},
  \bibinfo{pages}{81} (\bibinfo{year}{2003}).

\bibitem[{\citenamefont{Agra{\"\i}t et~al.}(1993)\citenamefont{Agra{\"\i}t,
  Rodrigo, and Vieira}}]{Agrait93}
\bibinfo{author}{\bibfnamefont{N.}~\bibnamefont{Agra{\"\i}t}},
  \bibinfo{author}{\bibfnamefont{J.~G.} \bibnamefont{Rodrigo}},
  \bibnamefont{and} \bibinfo{author}{\bibfnamefont{S.}~\bibnamefont{Vieira}},
  \bibinfo{journal}{Phys. Rev. B} \textbf{\bibinfo{volume}{47}},
  \bibinfo{pages}{12345} (\bibinfo{year}{1993}).

\bibitem[{\citenamefont{Brandbyge et~al.}(1995)\citenamefont{Brandbyge,
  Schi{\o}tz, S{\o}rensen, Stoltze, Jacobsen, N{\o}rskov, Olesen,
  L{\ae}gsgaard, Stensgaard, and Besenbacher}}]{Brandbyge95}
\bibinfo{author}{\bibfnamefont{M.}~\bibnamefont{Brandbyge}},
  \bibinfo{author}{\bibfnamefont{J.}~\bibnamefont{Schi{\o}tz}},
  \bibinfo{author}{\bibfnamefont{M.~R.} \bibnamefont{S{\o}rensen}},
  \bibinfo{author}{\bibfnamefont{P.}~\bibnamefont{Stoltze}},
  \bibinfo{author}{\bibfnamefont{K.~W.} \bibnamefont{Jacobsen}},
  \bibinfo{author}{\bibfnamefont{J.~K.} \bibnamefont{N{\o}rskov}},
  \bibinfo{author}{\bibfnamefont{L.}~\bibnamefont{Olesen}},
  \bibinfo{author}{\bibfnamefont{E.}~\bibnamefont{L{\ae}gsgaard}},
  \bibinfo{author}{\bibfnamefont{I.}~\bibnamefont{Stensgaard}},
  \bibnamefont{and}
  \bibinfo{author}{\bibfnamefont{F.}~\bibnamefont{Besenbacher}},
  \bibinfo{journal}{Phys. Rev. B} \textbf{\bibinfo{volume}{52}},
  \bibinfo{pages}{8499} (\bibinfo{year}{1995}).

\bibitem[{\citenamefont{Krans et~al.}(1995)\citenamefont{Krans, van Ruitenbeek,
  Fisun, Yanson, and de~Jongh}}]{Krans95}
\bibinfo{author}{\bibfnamefont{J.~M.} \bibnamefont{Krans}},
  \bibinfo{author}{\bibfnamefont{J.~M.} \bibnamefont{van Ruitenbeek}},
  \bibinfo{author}{\bibfnamefont{V.~V.} \bibnamefont{Fisun}},
  \bibinfo{author}{\bibfnamefont{I.~K.} \bibnamefont{Yanson}},
  \bibnamefont{and} \bibinfo{author}{\bibfnamefont{L.~J.}
  \bibnamefont{de~Jongh}}, \bibinfo{journal}{Nature}
  \textbf{\bibinfo{volume}{375}}, \bibinfo{pages}{767} (\bibinfo{year}{1995}).

\bibitem[{\citenamefont{Scheer et~al.}(1998)\citenamefont{Scheer,
  {Agra\"{\i}t}, Cuevas, {Levy Yeyati}, Ludoph, Mart{\'\i}n-Rodero, {Rubio
  Bollinger}, van Ruitenbeek, and Urbina}}]{Scheer98}
\bibinfo{author}{\bibfnamefont{E.}~\bibnamefont{Scheer}},
  \bibinfo{author}{\bibfnamefont{N.}~\bibnamefont{{Agra\"{\i}t}}},
  \bibinfo{author}{\bibfnamefont{J.~C.} \bibnamefont{Cuevas}},
  \bibinfo{author}{\bibfnamefont{A.}~\bibnamefont{{Levy Yeyati}}},
  \bibinfo{author}{\bibfnamefont{B.}~\bibnamefont{Ludoph}},
  \bibinfo{author}{\bibfnamefont{A.}~\bibnamefont{Mart{\'\i}n-Rodero}},
  \bibinfo{author}{\bibfnamefont{G.}~\bibnamefont{{Rubio Bollinger}}},
  \bibinfo{author}{\bibfnamefont{J.~M.} \bibnamefont{van Ruitenbeek}},
  \bibnamefont{and} \bibinfo{author}{\bibfnamefont{C.}~\bibnamefont{Urbina}},
  \bibinfo{journal}{Nature} \textbf{\bibinfo{volume}{394}},
  \bibinfo{pages}{154} (\bibinfo{year}{1998}).

\bibitem[{\citenamefont{van~den Brom and van Ruitenbeek}(1999)}]{Brom99}
\bibinfo{author}{\bibfnamefont{H.~E.} \bibnamefont{van~den Brom}}
  \bibnamefont{and} \bibinfo{author}{\bibfnamefont{J.~M.} \bibnamefont{van
  Ruitenbeek}}, \bibinfo{journal}{Phys. Rev. Lett.}
  \textbf{\bibinfo{volume}{82}}, \bibinfo{pages}{1526} (\bibinfo{year}{1999}).

\bibitem[{\citenamefont{Kondo and Takayanagi}(1997)}]{Kondo97}
\bibinfo{author}{\bibfnamefont{Y.}~\bibnamefont{Kondo}} \bibnamefont{and}
  \bibinfo{author}{\bibfnamefont{K.}~\bibnamefont{Takayanagi}},
  \bibinfo{journal}{Phys. Rev. Lett.} \textbf{\bibinfo{volume}{79}},
  \bibinfo{pages}{3455} (\bibinfo{year}{1997}).

\bibitem[{\citenamefont{Stafford et~al.}(1997)\citenamefont{Stafford,
  Baeriswyl, and B{\"u}rki}}]{Stafford97a}
\bibinfo{author}{\bibfnamefont{C.~A.} \bibnamefont{Stafford}},
  \bibinfo{author}{\bibfnamefont{D.}~\bibnamefont{Baeriswyl}},
  \bibnamefont{and}
  \bibinfo{author}{\bibfnamefont{J.}~\bibnamefont{B{\"u}rki}},
  \bibinfo{journal}{Phys. Rev. Lett.} \textbf{\bibinfo{volume}{79}},
  \bibinfo{pages}{2863} (\bibinfo{year}{1997}).

\bibitem[{\citenamefont{van Ruitenbeek et~al.}(1997)\citenamefont{van
  Ruitenbeek, Devoret, Esteve, and Urbina}}]{Ruitenbeek97}
\bibinfo{author}{\bibfnamefont{J.}~\bibnamefont{van Ruitenbeek}},
  \bibinfo{author}{\bibfnamefont{M.}~\bibnamefont{Devoret}},
  \bibinfo{author}{\bibfnamefont{D.}~\bibnamefont{Esteve}}, \bibnamefont{and}
  \bibinfo{author}{\bibfnamefont{C.}~\bibnamefont{Urbina}},
  \bibinfo{journal}{Phys. Rev. B} \textbf{\bibinfo{volume}{56}},
  \bibinfo{pages}{12566} (\bibinfo{year}{1997}).

\bibitem[{\citenamefont{Yannouleas et~al.}(1998)\citenamefont{Yannouleas,
  Bogachek, and Landman}}]{Yannouleas98}
\bibinfo{author}{\bibfnamefont{C.}~\bibnamefont{Yannouleas}},
  \bibinfo{author}{\bibfnamefont{E.}~\bibnamefont{Bogachek}}, \bibnamefont{and}
  \bibinfo{author}{\bibfnamefont{U.}~\bibnamefont{Landman}},
  \bibinfo{journal}{Phys. Rev. B} \textbf{\bibinfo{volume}{57}},
  \bibinfo{pages}{4872} (\bibinfo{year}{1998}).

\bibitem[{\citenamefont{Kassubek et~al.}(2001)\citenamefont{Kassubek, Stafford,
  Grabert, and Goldstein}}]{Kassubek01}
\bibinfo{author}{\bibfnamefont{F.}~\bibnamefont{Kassubek}},
  \bibinfo{author}{\bibfnamefont{C.~A.} \bibnamefont{Stafford}},
  \bibinfo{author}{\bibfnamefont{H.}~\bibnamefont{Grabert}}, \bibnamefont{and}
  \bibinfo{author}{\bibfnamefont{R.~E.} \bibnamefont{Goldstein}},
  \bibinfo{journal}{Nonlinearity} \textbf{\bibinfo{volume}{14}},
  \bibinfo{pages}{167} (\bibinfo{year}{2001}).

\bibitem[{\citenamefont{Zhang et~al.}(2003)\citenamefont{Zhang, Kassubek, and
  Stafford}}]{Zhang03}
\bibinfo{author}{\bibfnamefont{C.-H.} \bibnamefont{Zhang}},
  \bibinfo{author}{\bibfnamefont{F.}~\bibnamefont{Kassubek}}, \bibnamefont{and}
  \bibinfo{author}{\bibfnamefont{C.~A.} \bibnamefont{Stafford}},
  \bibinfo{journal}{Phys. Rev. B} \textbf{\bibinfo{volume}{68}},
  \bibinfo{pages}{165414} (\bibinfo{year}{2003}).

\bibitem[{\citenamefont{Urban and Grabert}(2003)}]{Urban03}
\bibinfo{author}{\bibfnamefont{D.~F.} \bibnamefont{Urban}} \bibnamefont{and}
  \bibinfo{author}{\bibfnamefont{H.}~\bibnamefont{Grabert}},
  \bibinfo{journal}{Phys. Rev. Lett.} \textbf{\bibinfo{volume}{91}},
  \bibinfo{pages}{256803} (\bibinfo{year}{2003}).

\bibitem[{\citenamefont{Urban et~al.}(2003)\citenamefont{Urban, B{\"u}rki,
  Zhang, Stafford, and Grabert}}]{Urban03b}
\bibinfo{author}{\bibfnamefont{D.~F.} \bibnamefont{Urban}},
  \bibinfo{author}{\bibfnamefont{J.}~\bibnamefont{B{\"u}rki}},
  \bibinfo{author}{\bibfnamefont{C.~H.} \bibnamefont{Zhang}},
  \bibinfo{author}{\bibfnamefont{C.~A.} \bibnamefont{Stafford}},
  \bibnamefont{and} \bibinfo{author}{\bibfnamefont{H.}~\bibnamefont{Grabert}},
  \bibinfo{howpublished}{cond-mat/0312517} (\bibinfo{year}{2003}).

\bibitem[{\citenamefont{Heer}(1993)}]{Heer93}
\bibinfo{author}{\bibfnamefont{W.~A.} \bibnamefont{Heer}},
  \bibinfo{journal}{Rev. Mod. Phys.} \textbf{\bibinfo{volume}{65}},
  \bibinfo{pages}{611} (\bibinfo{year}{1993}).

\bibitem[{\citenamefont{Yanson et~al.}(1999)\citenamefont{Yanson, Yanson, and
  van Ruitenbeek}}]{Yanson99}
\bibinfo{author}{\bibfnamefont{A.~I.} \bibnamefont{Yanson}},
  \bibinfo{author}{\bibfnamefont{I.~K.} \bibnamefont{Yanson}},
  \bibnamefont{and} \bibinfo{author}{\bibfnamefont{J.~M.} \bibnamefont{van
  Ruitenbeek}}, \bibinfo{journal}{Nature} \textbf{\bibinfo{volume}{400}},
  \bibinfo{pages}{144} (\bibinfo{year}{1999}).

\bibitem[{\citenamefont{Yanson et~al.}(2000)\citenamefont{Yanson, Yanson, and
  van Ruitenbeek}}]{Yanson00}
\bibinfo{author}{\bibfnamefont{A.~I.} \bibnamefont{Yanson}},
  \bibinfo{author}{\bibfnamefont{I.~K.} \bibnamefont{Yanson}},
  \bibnamefont{and} \bibinfo{author}{\bibfnamefont{J.~M.} \bibnamefont{van
  Ruitenbeek}}, \bibinfo{journal}{Phys. Rev. Lett.}
  \textbf{\bibinfo{volume}{84}}, \bibinfo{pages}{5832} (\bibinfo{year}{2000}).

\bibitem[{\citenamefont{Yanson et~al.}(2001)\citenamefont{Yanson, Yanson, and
  van Ruitenbeek}}]{Yanson01}
\bibinfo{author}{\bibfnamefont{A.~I.} \bibnamefont{Yanson}},
  \bibinfo{author}{\bibfnamefont{I.~K.} \bibnamefont{Yanson}},
  \bibnamefont{and} \bibinfo{author}{\bibfnamefont{J.~M.} \bibnamefont{van
  Ruitenbeek}}, \bibinfo{journal}{Fizika Nizkikh Temperatur}
  \textbf{\bibinfo{volume}{27}}, \bibinfo{pages}{1092} (\bibinfo{year}{2001}).

\bibitem[{\citenamefont{Landman et~al.}(1990)\citenamefont{Landman, Luedtke,
  Burnham, and Colton}}]{Landman90}
\bibinfo{author}{\bibfnamefont{U.}~\bibnamefont{Landman}},
  \bibinfo{author}{\bibfnamefont{W.~D.} \bibnamefont{Luedtke}},
  \bibinfo{author}{\bibfnamefont{N.~A.} \bibnamefont{Burnham}},
  \bibnamefont{and} \bibinfo{author}{\bibfnamefont{R.~J.}
  \bibnamefont{Colton}}, \bibinfo{journal}{Science}
  \textbf{\bibinfo{volume}{248}}, \bibinfo{pages}{454} (\bibinfo{year}{1990}).

\bibitem[{\citenamefont{Todorov and Sutton}(1996)}]{Todorov96}
\bibinfo{author}{\bibfnamefont{T.~N.} \bibnamefont{Todorov}} \bibnamefont{and}
  \bibinfo{author}{\bibfnamefont{A.~P.} \bibnamefont{Sutton}},
  \bibinfo{journal}{Phys. Rev. B} \textbf{\bibinfo{volume}{54}},
  \bibinfo{pages}{R14234} (\bibinfo{year}{1996}).

\bibitem[{\citenamefont{S{\o}rensen et~al.}(1998)\citenamefont{S{\o}rensen,
  Brandbyge, and Jacobsen}}]{Sorensen98}
\bibinfo{author}{\bibfnamefont{M.~R.} \bibnamefont{S{\o}rensen}},
  \bibinfo{author}{\bibfnamefont{M.}~\bibnamefont{Brandbyge}},
  \bibnamefont{and} \bibinfo{author}{\bibfnamefont{K.~W.}
  \bibnamefont{Jacobsen}}, \bibinfo{journal}{Phys. Rev. B}
  \textbf{\bibinfo{volume}{57}}, \bibinfo{pages}{3283} (\bibinfo{year}{1998}).

\bibitem[{\citenamefont{Brack}(1993)}]{Brack93}
\bibinfo{author}{\bibfnamefont{M.}~\bibnamefont{Brack}}, \bibinfo{journal}{Rev.
  Mod. Phys.} \textbf{\bibinfo{volume}{65}}, \bibinfo{pages}{677}
  (\bibinfo{year}{1993}).

\bibitem[{\citenamefont{Datta}(1995)}]{Datta95}
\bibinfo{author}{\bibfnamefont{S.}~\bibnamefont{Datta}},
  \emph{\bibinfo{title}{Electronic Transport in Mesoscopic Systems}}
  (\bibinfo{publisher}{Cambridge University Press}, \bibinfo{year}{1995}).

\bibitem[{\citenamefont{Torres et~al.}(1993)\citenamefont{Torres, Pascual, and
  S{\'a}enz}}]{Torres93}
\bibinfo{author}{\bibfnamefont{J.~A.} \bibnamefont{Torres}},
  \bibinfo{author}{\bibfnamefont{J.~I.} \bibnamefont{Pascual}},
  \bibnamefont{and} \bibinfo{author}{\bibfnamefont{J.~J.}
  \bibnamefont{S{\'a}enz}}, \bibinfo{journal}{Phys. Rev. B}
  \textbf{\bibinfo{volume}{49}}, \bibinfo{pages}{16581} (\bibinfo{year}{1993}).

\bibitem[{\citenamefont{B{\"u}rki et~al.}(1999)\citenamefont{B{\"u}rki,
  Stafford, Zotos, and Baeriswyl}}]{Buerki99a}
\bibinfo{author}{\bibfnamefont{J.}~\bibnamefont{B{\"u}rki}},
  \bibinfo{author}{\bibfnamefont{C.~A.} \bibnamefont{Stafford}},
  \bibinfo{author}{\bibfnamefont{X.}~\bibnamefont{Zotos}}, \bibnamefont{and}
  \bibinfo{author}{\bibfnamefont{D.}~\bibnamefont{Baeriswyl}},
  \bibinfo{journal}{Phys. Rev. B} \textbf{\bibinfo{volume}{60}},
  \bibinfo{pages}{5000} (\bibinfo{year}{1999}).

\bibitem[{\citenamefont{Kassubek et~al.}(1999)\citenamefont{Kassubek, Stafford,
  and Grabert}}]{Kassubek99}
\bibinfo{author}{\bibfnamefont{F.}~\bibnamefont{Kassubek}},
  \bibinfo{author}{\bibfnamefont{C.~A.} \bibnamefont{Stafford}},
  \bibnamefont{and} \bibinfo{author}{\bibfnamefont{H.}~\bibnamefont{Grabert}},
  \bibinfo{journal}{Phys. Rev. B} \textbf{\bibinfo{volume}{59}},
  \bibinfo{pages}{7560} (\bibinfo{year}{1999}).

\bibitem[{\citenamefont{B{\"u}rki and Stafford}(1999)}]{Buerki99b}
\bibinfo{author}{\bibfnamefont{J.}~\bibnamefont{B{\"u}rki}} \bibnamefont{and}
  \bibinfo{author}{\bibfnamefont{C.~A.} \bibnamefont{Stafford}},
  \bibinfo{journal}{Phys. Rev. Lett.} \textbf{\bibinfo{volume}{83}},
  \bibinfo{pages}{3342} (\bibinfo{year}{1999}).

\bibitem[{\citenamefont{Stafford et~al.}(1999)\citenamefont{Stafford, Kassubek,
  B{\"u}rki, and Grabert}}]{Stafford99}
\bibinfo{author}{\bibfnamefont{C.~A.} \bibnamefont{Stafford}},
  \bibinfo{author}{\bibfnamefont{F.}~\bibnamefont{Kassubek}},
  \bibinfo{author}{\bibfnamefont{J.}~\bibnamefont{B{\"u}rki}},
  \bibnamefont{and} \bibinfo{author}{\bibfnamefont{H.}~\bibnamefont{Grabert}},
  \bibinfo{journal}{Phys. Rev. Lett.} \textbf{\bibinfo{volume}{83}},
  \bibinfo{pages}{4836} (\bibinfo{year}{1999}).

\bibitem[{\citenamefont{Stafford et~al.}(2000)\citenamefont{Stafford, Kassubek,
  B{\"u}rki, Grabert, and Baeriswyl}}]{Stafford00}
\bibinfo{author}{\bibfnamefont{C.~A.} \bibnamefont{Stafford}},
  \bibinfo{author}{\bibfnamefont{F.}~\bibnamefont{Kassubek}},
  \bibinfo{author}{\bibfnamefont{J.}~\bibnamefont{B{\"u}rki}},
  \bibinfo{author}{\bibfnamefont{H.}~\bibnamefont{Grabert}}, \bibnamefont{and}
  \bibinfo{author}{\bibfnamefont{D.}~\bibnamefont{Baeriswyl}}, in
  \emph{\bibinfo{booktitle}{Quantum physics at the mesoscopic scale}}, edited
  by \bibinfo{editor}{\bibfnamefont{D.~C.} \bibnamefont{Glattli}},
  \bibinfo{editor}{\bibfnamefont{M.}~\bibnamefont{Sanquer}}, \bibnamefont{and}
  \bibinfo{editor}{\bibfnamefont{J.}~\bibnamefont{Tran~Thanh}}
  (\bibinfo{publisher}{EDP Sciences}, \bibinfo{address}{Les Ulis, France},
  \bibinfo{year}{2000}), pp. \bibinfo{pages}{49--53}.

\bibitem[{\citenamefont{Chandrasekhar}(1981)}]{Cha81}
\bibinfo{author}{\bibfnamefont{S.}~\bibnamefont{Chandrasekhar}},
  \emph{\bibinfo{title}{Hydrodynamic and Hydromagnetic Stability}}
  (\bibinfo{publisher}{Dover, New York}, \bibinfo{year}{1981}).

\bibitem[{\citenamefont{Tyson and Miller}(1977)}]{Tyson77}
\bibinfo{author}{\bibfnamefont{W.~R.} \bibnamefont{Tyson}} \bibnamefont{and}
  \bibinfo{author}{\bibfnamefont{W.~A.} \bibnamefont{Miller}},
  \bibinfo{journal}{Surf. Sci.} \textbf{\bibinfo{volume}{62}},
  \bibinfo{pages}{267} (\bibinfo{year}{1977}).

\bibitem[{\citenamefont{Brack and Bhaduri}(1997)}]{Brack97}
\bibinfo{author}{\bibfnamefont{M.}~\bibnamefont{Brack}} \bibnamefont{and}
  \bibinfo{author}{\bibfnamefont{R.~K.} \bibnamefont{Bhaduri}},
  \emph{\bibinfo{title}{Semiclassical Physics}}, vol.~\bibinfo{volume}{96} of
  \emph{\bibinfo{series}{Frontiers in Physics}} (\bibinfo{publisher}{Addison
  Wesley}, \bibinfo{year}{1997}).

\bibitem[{\citenamefont{Gutzwiller}(1997)}]{Gutzwiller90}
\bibinfo{author}{\bibfnamefont{M.~C.} \bibnamefont{Gutzwiller}},
  \emph{\bibinfo{title}{Chaos in classical and quantum mechanicanics}}
  (\bibinfo{publisher}{Addison-Wesley}, \bibinfo{address}{Reading (Mass.)},
  \bibinfo{year}{1997}).

\bibitem[{\citenamefont{B{\"u}rki and Stafford}(2001)}]{Buerki01}
\bibinfo{author}{\bibfnamefont{J.}~\bibnamefont{B{\"u}rki}} \bibnamefont{and}
  \bibinfo{author}{\bibfnamefont{C.~A.} \bibnamefont{Stafford}}, in
  \emph{\bibinfo{booktitle}{Electronic Correlations: From Meso- to
  Nano-Physics}}, edited by
  \bibinfo{editor}{\bibfnamefont{T.}~\bibnamefont{Martin}},
  \bibinfo{editor}{\bibfnamefont{G.}~\bibnamefont{Montambaux}},
  \bibnamefont{and} \bibinfo{editor}{\bibfnamefont{J.}~\bibnamefont{Tran
  Thanh~Van}} (\bibinfo{publisher}{EDP Sciences}, \bibinfo{address}{Les Ulis,
  France}, \bibinfo{year}{2001}), pp. \bibinfo{pages}{27--30}.

\bibitem[{\citenamefont{Puska et~al.}(2001)\citenamefont{Puska, Ogando, and
  Zabala}}]{Puska01}
\bibinfo{author}{\bibfnamefont{M.~J.} \bibnamefont{Puska}},
  \bibinfo{author}{\bibfnamefont{E.}~\bibnamefont{Ogando}}, \bibnamefont{and}
  \bibinfo{author}{\bibfnamefont{N.}~\bibnamefont{Zabala}},
  \bibinfo{journal}{Phys. Rev. B} \textbf{\bibinfo{volume}{64}},
  \bibinfo{pages}{033401} (\bibinfo{year}{2001}).

\bibitem[{\citenamefont{Ogando et~al.}(2002)\citenamefont{Ogando, Zabala, and
  Puska}}]{Ogando02}
\bibinfo{author}{\bibfnamefont{E.}~\bibnamefont{Ogando}},
  \bibinfo{author}{\bibfnamefont{N.}~\bibnamefont{Zabala}}, \bibnamefont{and}
  \bibinfo{author}{\bibfnamefont{M.~J.} \bibnamefont{Puska}},
  \bibinfo{journal}{Nanotechnology} \textbf{\bibinfo{volume}{13}},
  \bibinfo{pages}{363} (\bibinfo{year}{2002}).

\bibitem[{\citenamefont{Rubio et~al.}(1996)\citenamefont{Rubio, Agra{\"\i}t,
  and Vieira}}]{Rubio96}
\bibinfo{author}{\bibfnamefont{G.}~\bibnamefont{Rubio}},
  \bibinfo{author}{\bibfnamefont{N.}~\bibnamefont{Agra{\"\i}t}},
  \bibnamefont{and} \bibinfo{author}{\bibfnamefont{S.}~\bibnamefont{Vieira}},
  \bibinfo{journal}{Phys. Rev. Lett.} \textbf{\bibinfo{volume}{76}},
  \bibinfo{pages}{2302} (\bibinfo{year}{1996}).

\bibitem[{\citenamefont{Ludoph and van Ruitenbeek}(2000)}]{Ludoph00}
\bibinfo{author}{\bibfnamefont{B.}~\bibnamefont{Ludoph}} \bibnamefont{and}
  \bibinfo{author}{\bibfnamefont{J.~M.} \bibnamefont{van Ruitenbeek}},
  \bibinfo{journal}{Phys. Rev. B} \textbf{\bibinfo{volume}{61}},
  \bibinfo{pages}{2273} (\bibinfo{year}{2000}).

\bibitem[{\citenamefont{Ludoph and van Ruitenbeek}(1999)}]{Ludoph99}
\bibinfo{author}{\bibfnamefont{B.}~\bibnamefont{Ludoph}} \bibnamefont{and}
  \bibinfo{author}{\bibfnamefont{J.~M.} \bibnamefont{van Ruitenbeek}},
  \bibinfo{journal}{Phys. Rev. B} \textbf{\bibinfo{volume}{59}},
  \bibinfo{pages}{12290} (\bibinfo{year}{1999}).

\bibitem[{\citenamefont{B\"urki et~al.}({\natexlab{a}})\citenamefont{B\"urki,
  Urban, and Stafford}}]{Buerki04b}
\bibinfo{author}{\bibfnamefont{J.}~\bibnamefont{B\"urki}},
  \bibinfo{author}{\bibfnamefont{D.~F.} \bibnamefont{Urban}}, \bibnamefont{and}
  \bibinfo{author}{\bibfnamefont{C.~A.} \bibnamefont{Stafford}},
  \bibinfo{note}{in preparation}.

\bibitem[{\citenamefont{B\"urki et~al.}({\natexlab{b}})\citenamefont{B\"urki,
  Stafford, and Stein}}]{Buerki04}
\bibinfo{author}{\bibfnamefont{J.}~\bibnamefont{B\"urki}},
  \bibinfo{author}{\bibfnamefont{C.~A.} \bibnamefont{Stafford}},
  \bibnamefont{and} \bibinfo{author}{\bibfnamefont{D.~L.} \bibnamefont{Stein}},
  \eprint{cond-mat/0406374}.

\end{thebibliography}

\end{document}